\documentclass[article,nojss]{jss}

\usepackage{thumbpdf,lmodern}
\usepackage{orcidlink}

\usepackage{framed}

\usepackage{amsmath}
\usepackage{amssymb}
\usepackage{mathtools}
\usepackage{multirow}
\usepackage{dsfont}
\usepackage{algorithmic}
\usepackage{algorithm}
\def\spacingset#1{\renewcommand{\baselinestretch}%
{#1}\small\normalsize} \spacingset{1}
\mathchardef\mhyphen="2D
\def \dsR {\text{$\mathds{R}$}}
\let\proglang=\textsf

\makeatletter
\def\@codex#1{{\normalfont\ttfamily\hyphenchar\font=-1 #1}\egroup}
\makeatother

\def \dsP {\text{$\mathds{P}$}}

\def \dsR {\text{$\mathds{R}$}}

    \def \mN {\text{\boldmath$N$}}

\def \xvec {\text{\boldmath$x$}}    \def \mX {\text{\boldmath$X$}}
\def \yvec {\text{\boldmath$y$}}

\def \betavec         {\text{\boldmath$\beta$}}

\def \omegavec        {\text{\boldmath$\omega$}}

\def \mOmega   {\mathbf{\Omega}}

\author{Nikolaus Umlauf~\orcidlink{0000-0003-2160-9803}\\Universit\"at Innsbruck
   \And Nadja Klein~\orcidlink{0000-0002-5072-5347}\\Humboldt-Universit\"at zu Berlin
}
\Plainauthor{Umlauf~N., Klein~N.}

\title{Distributional Adaptive Soft Regression Trees}
\Plaintitle{Distributional Adaptive Soft Regression Trees}
\Shorttitle{Distributional Adaptive Soft Regression Trees}

\Abstract{
  Random forests are an ensemble method relevant for many problems, such as regression or
  classification. They are popular due to their good predictive performance (compared to, e.g.,
  decision trees) requiring only minimal tuning of hyperparameters. They are built via
  aggregation of multiple regression trees during training and are usually calculated recursively
  using hard splitting rules. Recently regression forests have been incorporated into the framework
  of distributional regression, a nowadays popular regression approach aiming at estimating
  complete conditional distributions rather than relating the mean of an output variable to input
  features only -- as done classically. This article proposes a new type of a distributional
  regression tree using a multivariate soft split rule. One great advantage of the soft split is
  that smooth high-dimensional functions can be estimated with only one tree while the complexity
  of the function is controlled adaptive by information criteria. Moreover, the search for the
  optimal split variable is obsolete. We show by means of extensive simulation studies
  that the algorithm has excellent properties and outperforms various benchmark methods,
  especially in the presence of complex non-linear feature interactions.
  Finally, we illustrate the usefulness of our approach with an example on probabilistic
  forecasts for the Sun's activity.
}

\Plainkeywords{Adaptive soft split; decision trees; generalized additive models for location, scale and shape; probabilistic forecasting, random forests}

\Address{
  Nikolaus Umlauf\\
  Department of Statistics\\
  Faculty of Economics and Statistics\\
  Universit\"at Innsbruck\\
  Universit\"atsstr.~15\\
  6020 Innsbruck, Austria\\
  E-mail: \email{Nikolaus.Umlauf@uibk.ac.at},\\
  URL: \url{https://eeecon.uibk.ac.at/~umlauf/},\\

  Nadja Klein\\
  Humboldt Universit\"at zu Berlin\\
  School of Business and Economics\\
  Chair of Statistics and Data Science\\
  Unter den Linden 6\\
  10099 Berlin, Germany\\
  E-mail: \email{nadja.klein@hu-berlin.de}\\
  URL: \url{https://hu.berlin/NK}
}

\begin{document}

\section{Introduction} \label{sec:intro}

In many applications it is important to model not only the expected value of a response variable,
but rather a complete probabilistic prediction model. While there have been proposed numerous methods to obtain such an also called \emph{distributional model} \citep[see e.g.~][for a recent review]{KneSaeSil2021}, in this paper we focus on the class of
 structured additive distributional regression \citep{Klein+Kneib+Klasen+Lang:2015} aka generalized additive model for location, scale and shape 
\citep[GAMLSS;][]{Rigby+Stasinopoulos:2005}, which can model every parameter of an arbitrary parametric target
distribution through input features thereby implicitly constituting  a probabilistic prediction model. This model class has been successfully used  in a number of applications to develop state-of-the-art
probabilistic forecasts \citep[see, e.g.,][]{Klein+Denuit+Kneib+Lang:2014,Simon+Fabsic+Mayr+Umlauf+Zeileis:2018, Simon+Mayr+Umlauf+Zeileis:2019,RConsort2020,Ser2011,Zie2021}.
In GAMLSS, each distribution parameter is typically modeled
with a structured additive predictor \citep[STAR;][] {Fahrmeir+Kneib+Lang+Marx:2013,Woo2017}
decomposed from a sum of smooth generic functions, following the structure of generalized additive models
\citep[GAM;][]{Hastie+Tibshirani:1990}. Although this class of models can already model fairly complex 
interactions, the ability to estimate higher dimensional functions with, for example,
more than three features using (tensor product) splines is very difficult and costly, since, for
example, the number of parameters to be estimated increases exponentially with each additional input variable.
To address this issue, distribution trees and forests have recently been proposed by 
\citet{distforest:paper}, which are based on split methods similar to the very established random forests
\citep[RF]{Breiman:2001}. In a similar vein, \citet{RueKolKle2021} propose to extend the STAR predictors of each distributional parameter by a deep neural network to allow for high-dimensional interaction terms and where identification is realized through a generic orthogonalization cell. Although RFs are known for their high flexibility, the splitting method
is still a limiting factor in modeling smooth high-dimensional functions. The reason is relatively
simple: a single regression tree is usually built using hard
splits, which lead to a tree modeling a step function. 
This in turn means that an RF, which consists of a number of trees, will in the best case find
only a ``nearly smooth function''. The step function approximation problem can sometimes be exacerbated
when there are larger gaps in the data and many covariates. Although \citet{Breiman:2001} provides an
upper bound on the generalization error and \citet{JMLR:v18:17-269} more recently show empirically and 
theoretically that increasing the size of an RF is beneficial, the approximation error as a 
consequence of the splitting rule is not considered. \citet{Ciampi:2002} first addressed this problem by 
instead using  a soft probabilistic splitting rule for growing a tree. This idea is followed to the
best of our knowledge only by few contributions
\citep[e.g.][]{10.1007/978-3-7908-1777-5_6, Frosst:2017, Linero:2018, Luo:2021}; but none of them considers a  distributional framework.
The prominence of a hard splitting rule is most likely due to the fact that a single tree yields a 
completely interpretable model and its robustness in mean regression applications. However, especially for
distributional models RFs are usually used as predictive models, where interpretation 
only plays a minor role. In \citet{irsoy2012soft} a multivariate version of the soft decision
tree is introduced and \citet{Yildiz2016} show excellent predictive performance when bagging multivariate soft
decision trees.

In this paper, we extend these ideas of soft splitting rules for the class of GAMLSS and propose a distributional adaptive soft regression tree (DAdaSoRT) for full
probabilistic forecasting that is capable of approximating both linear and complex non-linear rough and 
smooth functions. Our main contributions are:
\begin{itemize}
\item A single distributional soft tree is grown using multivariate soft splits, which has in most cases
  better, at least similar, approximation capabilities than a complete distributional forest.
\item The tree growth algorithm is adaptive and the optimal size is selected based on information
  criteria such as the Akaike information criterion (AIC) or Bayesian information criterion (BIC), which lead to final trees with relatively few degrees of freedom.
\item To further avoid the problem of overfitting, additional shrinkage parameters control
  the smoothness of the estimated functions and are the only hyperparameters.
  \item The efficacy of our proposed method is demonstrated in a benchmark study and in a forecast exercise for predicting the Sun's activity based on  time series data.
\end{itemize}

The remainder of this paper is organized as follows: In Section~\ref{sec:asofttree} we first introduce the 
ideas of classical soft regression trees in detail before presenting the full adaptive version of a soft tree. Then, in Section~\ref{sec:sdrt}, we extend the adaptive soft regression trees to distributional 
regression based on GAMLSS (referred to as DAdaSoRT) and provide algorithms for estimation. In Section~\ref{sec:properties} we present further 
properties and details of  distributional adaptive soft regression trees. In Section~\ref{sec:software}
we briefly introduce the accompanying \proglang{R} package \pkg{softtrees}.
An extensive simulation study that examines the performance of soft distributional regression trees
compared to classical GAMLSS models and distributional forests is presented in
Section~\ref{sec:simulation}. In Section~\ref{sec:application} we then show the benefits of the
approach using a complex problem of probabilistic forecasting the Sun's solar cycles 25 and 26.

\section{Adaptive Soft Regression Trees} \label{sec:asofttree}

In this section, we first review the classical multivariate soft regression trees before we introduce our
improved version which we call adaptive soft regression trees (AdaSoRT) and which allows to obtain more efficient predictions
through local adaptive smoothing in the model fits.

\subsection{Classical Soft Regression Trees}\label{sec:srt}
Classical regression trees use hard binary splits as a tree grows. The exact split point is searched in
different ways, e.g., by an exhaustive search or by decision rules based on instability tests
\citep{Hothorn+Hornik+Zeileis:2006}. In contrast to this, soft regression trees (SoRT) are grown with so-called soft
splitting rules, which means that instead of a step function a soft discriminator is used, a smooth continuous function
that does not have an exact split point.
Each terminal node is influenced by all
observation points through the weighting created in this way.

\begin{figure}[!t]
\centering
\includegraphics[width=0.5\textwidth]{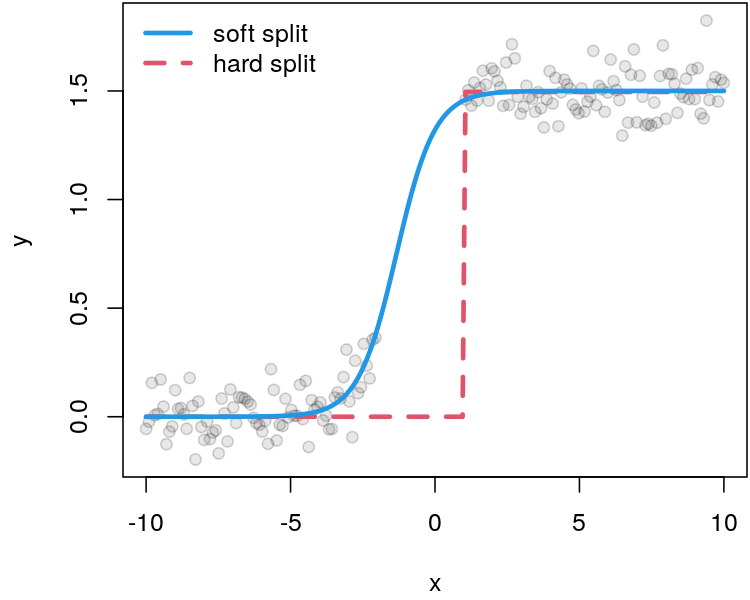}
\caption{\label{fig:splits} Illustrating example. Shown are bivariate simulated data $(y,x)$ (light-grey points) together with a  classical hard split (dashed red line) and a
  soft split using a logistic function (blue solid line).}
\end{figure}

An example that illustrates this difference along simulated bivariate data is shown in Figure~\ref{fig:splits}. Here, the hard
binary split (dashed red line) separates the data at $x = 1$ and yields only two possible predictions, i.e.~$\hat y=0$ for $x<1$ and $\hat y=1.5$ for $x\geq 1$. Instead, the soft split (blue line) allows a smooth
transition. Intuitively, rather than assigning observations to single nodes, a soft split uses a better balanced weighting of all possible nodes.

The literature differentiates between univariate and multivariate soft splits. While univariate soft splits \citep[as, e.g., introduced by][for binary outcomes]{Ciampi:2002} typically outperform hard splits, they still require finding the optimum split variable, which can be computationally intensive, in particular when the feature space is large.
To reduce the computational burden, \citet{irsoy2012soft} instead proposed a multivariate soft node which is in principal similar
to the nodes of a neural network (NN). In that respect, \citet{Frosst:2017} distil a NN into a soft decision tree to improve upon  NN classification tasks and \citet{Luo:2021} show that soft decision trees can be
an alternative to deep NNs for numerous regression tasks.

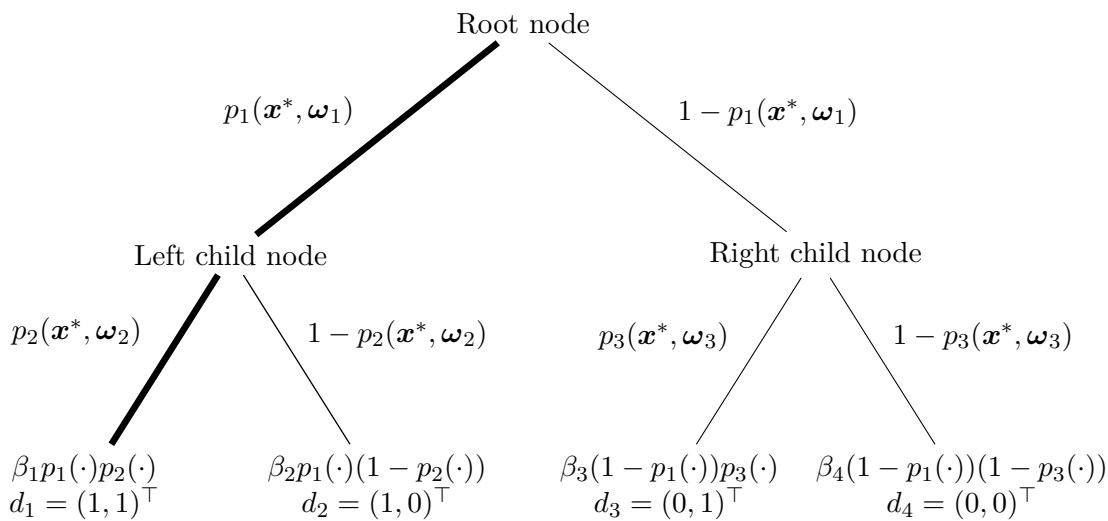
\begin{figure}[ht!]
\centering
\begin{tikzpicture}[sibling distance=10em, level distance=8em,
  every node/.style = {shape=rectangle, rounded corners,
    draw=none, align=center,
    top color=white},
  ered/.style={edge from parent/.style={red,very thick,draw}}]
\node{ Root node} 
    child    { node { Left child node} 
        child { node {$\beta_1p_1(\cdot)p_2(\cdot)$\\$d_1 = (1, 1)^\top$}
        edge from parent [black, line width=2.5pt] node[above left] { \color{black}{$p_2(\xvec^\ast, \omegavec_2)$} }}
        child { node {$\beta_2p_1(\cdot)(1-p_2(\cdot))$\\$d_2 = (1, 0)^\top$}
        edge from parent [black, line width=0.5pt] node[above right] { \color{black}{$1 - p_2(\xvec^\ast, \omegavec_2)$} }}
        edge from parent [black, line width=2.5pt] node[above left] { \color{black}{$p_1(\xvec^\ast, \omegavec_1)$} }
    }
    child [missing]
    child { node { Right child node}
        child { node {$\beta_3(1-p_1(\cdot))p_3(\cdot)$\\$d_3 = (0, 1)^\top$}
        edge from parent node[above left] { $p_3(\xvec^\ast, \omegavec_3)$ }}
        child { node {$\beta_4(1-p_1(\cdot))(1-p_3(\cdot))$\\$d_4 = (0, 0)^\top$}
        edge from parent node[above right] { $1 - p_3(\xvec^\ast, \omegavec_3)$ }}
        edge from parent node[above right] { $1 - p_1(\xvec^\ast, \omegavec_1)$ }
    };
\end{tikzpicture}
\caption{\label{fig:softtree} Illustrating tree structure. A SoRT with $4$ terminal nodes.}
\end{figure}

Now suppose there is data $\yvec=(y_1,\ldots,y_n)^\top$, such that for each output $y_i$, $i=1,\ldots,n$ there is a $q$-dimensional feature vector
$\xvec_i = (x_{i1}, \ldots, x_{iq})^\top$ available. Let furthermore 
$\mathcal{N}=	\mathcal{M} \,\dot\cup\, \mathcal{T}$ be the set of all nodes (excluding the root node)
consisting of the disjoint union of the set of all nodes that have children (i.e., 
those that serve as root nodes for some others) $\mathcal{M}$ and the set of terminal 
(leaf) nodes $\mathcal{T}$, such that $|\mathcal{M}|=M$, $|\mathcal{T}|=T$ and 
$|\mathcal{N}|=M+T$.  For growing a SoRT
any root node $l\in\mathcal{M}$, with output $N_l( \cdot )$ is ``split 
softly'' into a weighted average of left and right child nodes, $N_l^L( \cdot )$ and $N_l^R( \cdot )$, respectively, where
\begin{equation} \label{eqn:softnode}
N_l(\xvec_i) = N_{l}^L(\xvec_i) \cdot p_l(\xvec_i) +
  N_{l}^R(\xvec_i) \cdot (1 - p_l(\xvec_i)),
\end{equation}
and the weighting function  $p_l(\cdot) \in [0, 1]$ can be seen as the posterior probability 
$\dsP(L\mid\xvec_i)=p_l(\xvec_i)$  of redirecting $y_i$  to the left child node given 
$\xvec_i$; and $\dsP(R\mid\xvec_i)=1-p_l(\xvec_i)$ being the corresponding probability for assignment to the right node.
A common choice for the mappings $p_l( \cdot ):\dsR\mapsto[0,1]$ is the sigmoid (logistic) function given by
\begin{equation} \label{eqn:logistic}
p_l(\xvec_i) = \frac{1}{1 + \exp(-(\xvec_i^\top\omegavec_l))},
\end{equation}
where $\omegavec_l$ are weights that need to be estimated from the data.
If $N_{l}^L(\xvec_i)$ and $N_{l}^R(\xvec_i)$ are terminal nodes, i.e.~$l\in\mathcal{T}$,
we can write
\begin{equation} \label{eqn:terminalnode}
N_l(\xvec_i) = \beta_{l}^L \cdot p_l(\xvec_i) +
  \beta_{l}^R \cdot (1 - p_l(\xvec_i)),
\end{equation}
with parameters $N_{l}^L(\xvec_i) = \beta_{l}^L$ and $N_{l}^R(\xvec_i) = \beta_{l}^R$ that
need to be estimated, too. This means, in each
SoRT growing step one of the nodes $N_{l}^L(\xvec_i)$ or
$N_{l}^R(\xvec_i)$ is replaced by another soft split node~\eqref{eqn:softnode} unless $l\in\mathcal{T}$.
Therefore, the multiplicative soft weighting that is enforced leads to a type of basis 
functions that adapt to the data in a smooth way, since at each node the function is differentiable
if and only if $p_l( \cdot )$ is differentiable. Finally, given the sets
$\mOmega_1,\ldots,\mOmega_T$ of weights involved in computing each of the $T$ terminal nodes as
well as $\mOmega$ denoting the set of all weights in the tree, 
predictions $\hat y^\ast$ from the final SoRT for any new feature $\xvec^\ast$ can be
computed by the linear combination 
$\mN(\xvec^\ast, \mOmega)^\top\boldsymbol{\beta}$, where $\betavec=(\beta_1,\ldots,\beta_T)^\top$
is the vector of terminal node parameters and
$\mN(\xvec^\ast, \mOmega)^\top=(P_1(\xvec^\ast,\mOmega_1),\ldots,P_T(\xvec^\ast,\mOmega_T))^\top$
is the $T$-dimensional design vector with path probabilities for the $T$ paths in the tree. More
precisely, let $D_l$ denote a path of length $\log_2(T)$ associated with
$P_l(\xvec^\ast,\mOmega_l)$. Let furthermore  $\mathcal{D}_l$ be the set of nodes involved in that
path $D_l$. Then, we can write
\begin{equation} \label{eqn:designcolumn}
P_{l}(\xvec^\ast, \mOmega_l) =
  \prod_{r \in \mathcal{D}_l} p_r(\xvec^\ast)^{d_r}(1 -
    p_r(\xvec^\ast))^{1 - d_r},
\end{equation}
with  $d_r \in \{0, 1\}$ indicating the binary directions (left/right) in each split along the path.
In Figure~\ref{fig:softtree} the construction of a SoRT is illustrated. Here, each path $D_l$, $l=1,\ldots,4$ from the top root node to one of the four terminal nodes
represents one column of the design matrix $\mN(\mX,\mOmega)\in\dsR^{n\times T}$, which are constructed by
multiplication of level specific probabilities as defined in \eqref{eqn:designcolumn}, and where 
$\mX=(\xvec_1|\ldots|\xvec_n)^\top$ is the $n\times q$ feature matrix.
For instance, the most left path $D_1$ (thick black lines in Figure~\ref{fig:softtree})
from $p_1$ to $p_2$ to $\beta_1p_1p_2$ has length 2. Its involved weights are
$\mOmega_1=\lbrace\omegavec_1,\omegavec_2\rbrace$.

\begin{figure}[!t]
\centering
\includegraphics[width=1\textwidth]{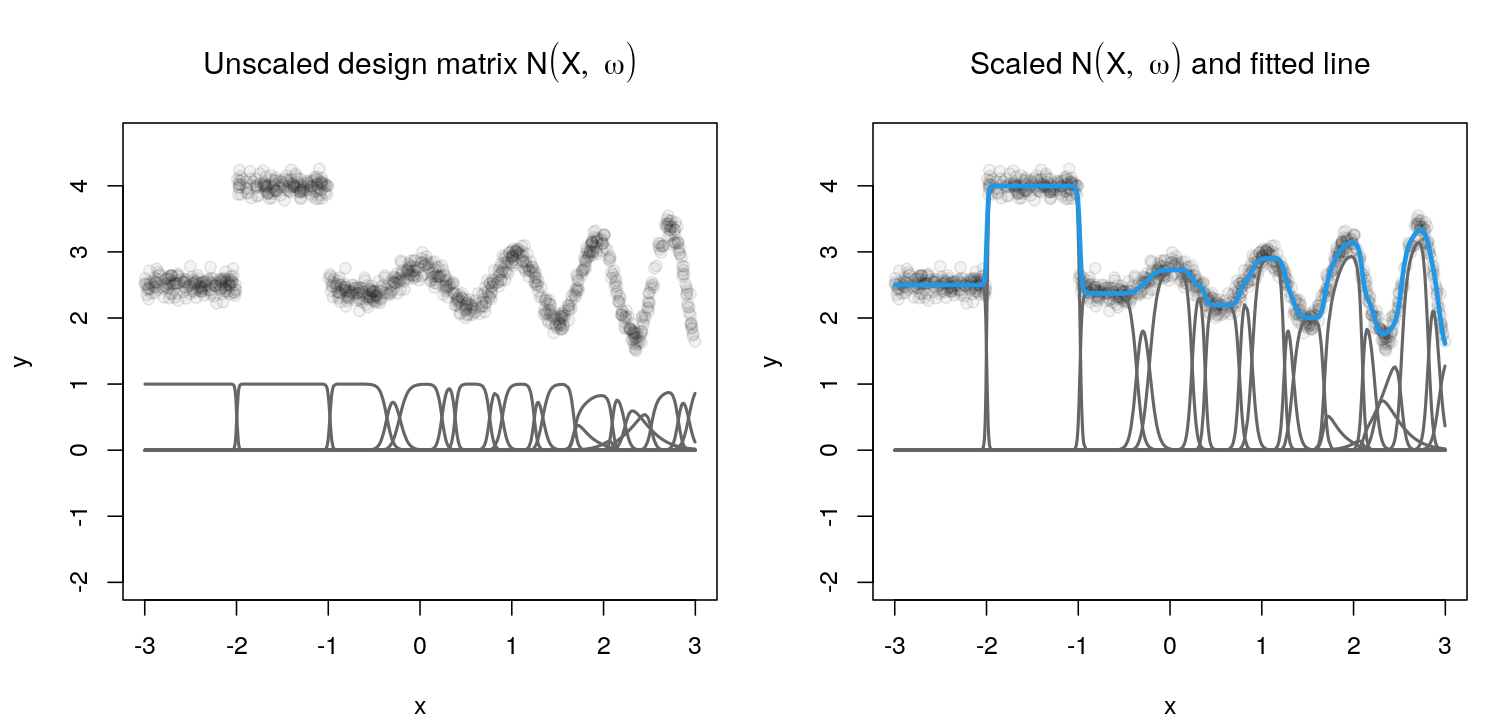}
\caption{\label{fig:design} Learned design matrix $\mN(\mX,\mOmega)$ using
  an univariate SoRT and training data. The left plot depits the data
  (light-grey points) together with the unscaled basis functions (solid grey lines).
  The right plot shows the basis functions 
  scaled with regression parameters $\boldsymbol{\beta}$ 
  together with the fitted curve $\mN(\mX,\mOmega)\betavec$ (blue line).}
\end{figure}

An illustration of the recursion for an $n$ dimensional training data set with feature matrix
$\mX$ and designmatrix $\mN(\mX,\mOmega)$ is
given in Figure \ref{fig:design}.
This figure shows  a step function
with subsequent highly oscillatory relationship between $x$ and $y$. The example shows very well
how a SoRT is suitable for both abrupt changes and smooth functional shapes.
Moreover, the SoRT can be
viewed as a method that learns basis functions, e.g., similar to B-spline basis functions used
for P-splines \citep{Eilers+Marx:1996}. The main difference to the latter is that the basis functions from
a SoRT are adaptive and smooth and can model very complex relationships, while the
number of terminal nodes (columns in the design matrix $\mN(\mX, \mOmega)$)
is much less compared to the degrees of freedom that are needed to obtain a similar fit with
(P)-splines. In this example, the tree only needs $T = 20$ terminal nodes to approximate the training data
quite well already and the corresponding spline model fitted with the \proglang{R} package
\texttt{mgcv} \citep{Wood:2021} requires $\approx 156$ degrees of freedom.

\begin{figure}[!t]
\centering
\includegraphics[width=1\textwidth]{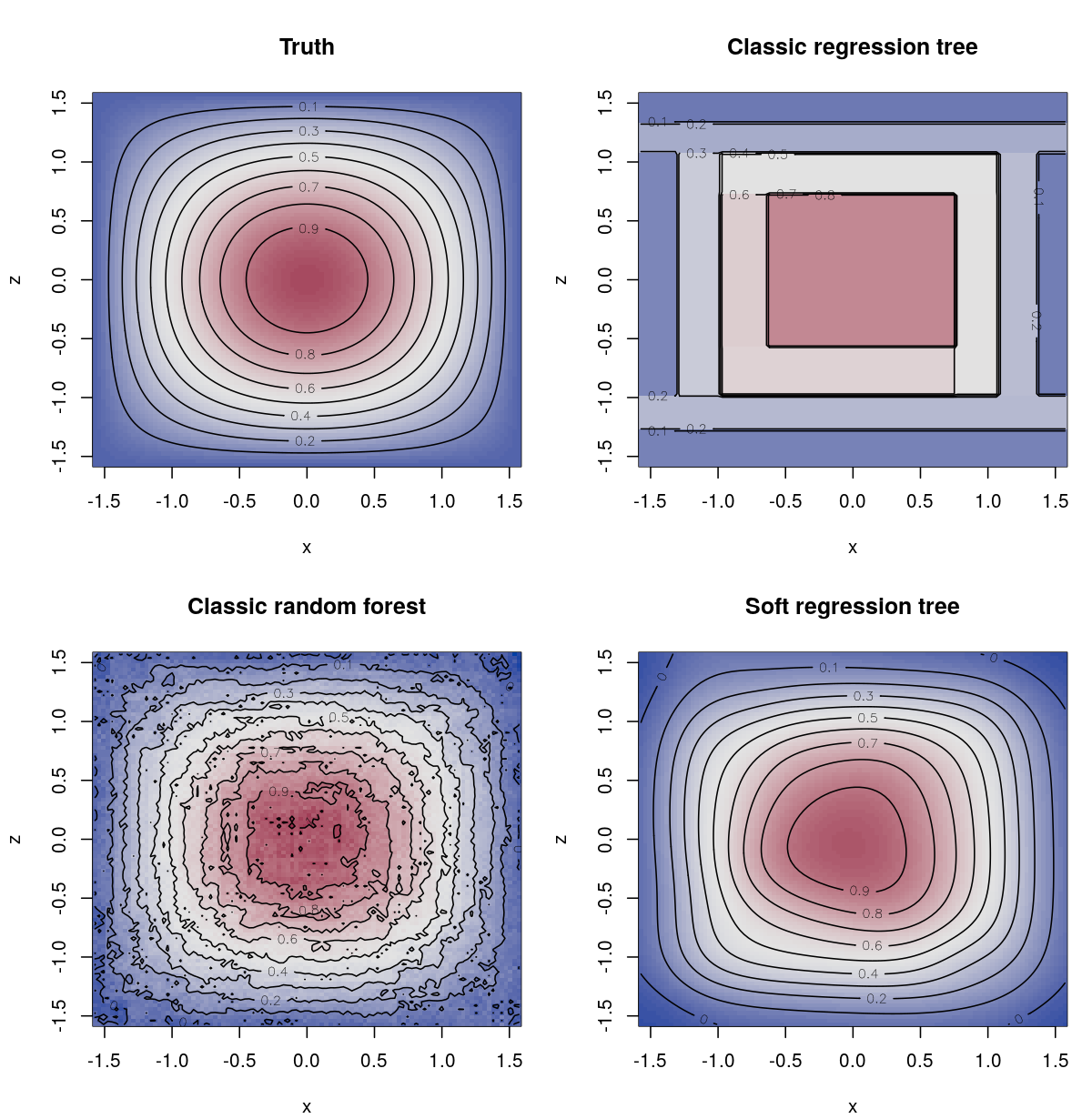}
\caption{\label{fig:circular} Artificial 2D data example. The true function is shown in the top
  left panel. The estimated function using a classic regression tree is shown in the top right panel,
  the estimated function using a RF with $2000$ trees is shown in the bottom left panel
  and the estimated function from the SoRT in the bottom right panel.}
\end{figure}
Another feature of the SoRT is its ability to model smooth interactions, such as those
often found in spatial regression problems. An example is given in 
Figure~\ref{fig:circular}.
Here, the true function is given by $f(x, z) = \sin(x) \cdot \sin(z)$ and
$10000$ data points are simulated using the Gaussian mean regression model
$y_i = f(x_i, z_i) + \varepsilon_i$ with $\varepsilon_i \sim N(0, 0.1^2)$ and based on 100
equidistant values $x,z \in [-\pi/2, \pi/2]$  each. The true function $f$ is given in the upper
left panel of Figure \ref{fig:circular}. The classical regression tree in the
upper right panel shows that the true function can only be very roughly approximated, which is
natural since the hard splits generated can only model rectangular regions. Therefore, to better
capture complex interactions, a RF is estimated 
in many regression situations \citep{Breiman:2001}. By growing several trees on data subsets
and then aggregating over the 
trees to predict the response, a RF is also able to model almost smooth transitions.
However, as can be seen in the lower left panel of Figure \ref{fig:circular} even a RF
with $2000$ trees is not able to reproduce the smooth surface of the true function very nicely,
as the shown contour lines depict a rather rough estimate. In contrast, the SoRT
with only 24 terminal nodes is able to reproduce the true functional form quite well
(compare the lower right panel). What we have now illustrated for 1-2 dimensions only naturally extends to much higher dimensional interactions making a SoRT particularly attractive in many data situations.

Technically, the reason for the good approximation abilities is mainly  the structural form of the 
SoRT, which is quite similar to a NN with a single hidden layer, which is known to be
a universal function approximator \citep[][and also Section \ref{sec:universalapproximator}]{Hornik:1991}.
However, as mentioned earlier, the SoRT iteratively learns the shape of the design matrix $\mN(\mX,\mOmega)$ in
multiple directions, whereas a NN with a single 
hidden layer is just a sum of scaled activation functions. In summary, the SoRT is a
very flexible tool for modeling complex relationships, it can represent both soft transitions and step
functions of a classical regression tree.

\subsection{Adaptive Soft Regression Tree Predictor}\label{sec:modification}

We will extend the basic soft tree structure of the previous Subsection~\ref{sec:srt} to allow for more efficient and better predictions.
For the moment keep in mind the example of a SoRT presented in Figure~\ref{fig:softtree}.
The basic idea of growing a tree is that we only allow new nodes to emerge at positions where there is still a lot of 
unexplained information. E.g., in Figure~\ref{fig:design}, an additional soft split in the range
$-3 < x < -1$ would not improve the model fit much, since the first split could already approximate the step
function behaviour quite well. Therefore, if there is a global functional shape that can be well
approximated with a few splits then technically it will be better to maintain the model fit in that region of the 
data, rather than always computing new models that include all terminal nodes. This means that growing the SoRT
should coarsely adapt to the  global functional form at the beginning and only model fine-grained information in
later steps as the tree grows. This behavior can be easily implemented by providing parameters to all $J=M+T$ nodes (except the root node)
of the SoRT rather than just to the terminal leaf nodes as done for classical SoRT \citep{Luo:2021}. To set this up, we do not only assign path probabilities of the form \eqref{eqn:designcolumn} to the terminal nodes but rather to all  nodes and call this version ``adaptive soft regression tree'' (AdaSoRT). Let thus $D_{(l)}$ be any path  from the root node to any node $N_l(\xvec)$. Let also similar to before  $\mathcal{D}_{(l)}$ be the set of nodes involved in forming in path $D_{(l)}$ of length $1\leq |\mathcal{D}_{(l)}|\leq\log_2(T)$, with path probability $P_{(l)}$ from \eqref{eqn:designcolumn} and set of weights $\mOmega_{(l)}$. With these  probabilities we then define a feature-specific
predictor of the form
\begin{equation} \label{eqn:mtree}
\eta(\xvec) = \beta_0 + \sum_{j = 1}^{J} P_j(\xvec, \mOmega_{(j)}) \beta_j,
\end{equation}
where $\beta_0$ is an overall intercept representing the root node.
Hence, for instance for the tree in Figure~\ref{fig:softtree}, we obtain $J =6$ elements for the  AdaSoRT or more generally $J=\sum_{l=1}^{\log_2(T)}2^l$. This tree structure thus allows 
us to decompose the predictor into coarse and fine elements, with the finer structures 
tending to be controlled by the rearmost elements of the sum of \eqref{eqn:mtree},
i.e., the terminal nodes in Figure~\ref{fig:softtree}.
The final AdaSoRT can again be represented by
$$
\boldsymbol{\eta} = \mN(\mX, \mOmega)\betavec,
$$
but now with $1 + M + T = 1+J$ columns in the design matrix $\mN( \cdot )$ (as opposed to $T$ columns for the classical SoRT),
 corresponding to the intercept and respective nodes that are generated in the growing steps and $\mOmega$ is now the set of
the weights for all paths together.

\section{AdaSoRT for Structured Additive Distributional Regression} \label{sec:sdrt}

SoRT introduced so far are  able to efficiently make point predictions  even when the underlying relation to covariates is rather  complex or high-dimensional. With the extension AdaSoRT to the adaptive version of SoRT of the previous section, we can make estimation and accuracy of point predictions even more efficient.  However, often predicting the conditional first moment  of the quantity of interest (i.e., the expected value) is not enough. Rather, being able to infer  complete predictive distributions and to derive further quantities from it (such as scale, prediction intervals, certain quantiles or tail measures) becomes more and more relevant in many modern applications
(e.g., electricity price forecasting, \citealp{NOWOTARSKI20181548} or climate change, \citealp{Raisanen2022})
; including the empirical illustration in this paper on forecasting the Sun's solar cycle 25 and 26 in Section~\ref{sec:application}. To address this task, we propose AdaSoRT  for structured additive distributional regression -- which we refer to as \emph{Distrubtional} AdaSoRT (DAdaSoRT) and present how these flexible models can be fitted efficiently  and how to perform model choice.

\subsection{Model Specification} \label{sec:modelspec}

The idea of structured additive distributional regression
\citep[or GAMLSS;][]{Rigby+Stasinopoulos:2005, Klein+Kneib+Klasen+Lang:2015}
is to model all parameters of an arbitrary parametric response distribution (rather than just the mean of an exponential family distribution) through available features.  Specifically, assume
\begin{equation} \label{eqn:dreg}
y \sim {{D}_y}\left(h_{1}(\theta_{1}) = \eta_{1}, \,\,
  h_{2}(\theta_{2}) = \eta_{2}, \dots,  h_{K}(\theta_{K}) =
  \eta_{K}\right),
\end{equation}
where ${{D}_y}$ denotes a parametric distribution for the response
variable $y$ \citep[which can in our framework also be non-continous or multivariate; see e.g.][]{Klein+Kneib+Lang:2015,Klein+Kneib+Klasen+Lang:2015} with $K$ parameters $\theta_k\equiv \theta_k(\xvec)$, $k = 1, \ldots, K$, that are linked to
feature-dependent predictors $\eta_{k}\equiv\eta_k(\xvec)$ using known monotonic and twice differentiable functions
$h_{k}(\cdot)$. The latter are simply chosen to meet potential parameter space restrictions on $\theta_k$.  The distribution $\mathcal{D}_y$ is arbitrary, but we assume it has a parametric probability density/mass function $f_y(\cdot;\theta_1,\ldots,\theta_k)$ and that this density is twice continuously differentiable with respect to all distributional predictors~$\eta_k$.

Using the AdaSoRT presented in Section~\ref{sec:modification},  the
$k$-th predictor for distributional parameter $\theta_k$ is then given by
\begin{eqnarray} \label{eqn:addpred}
\boldsymbol{\eta}_k \equiv \eta_k(\mX_k; \boldsymbol{\beta}_k, \mOmega_{k}) &=&
  \mN_k(\mX_k, \mOmega_{k})\boldsymbol{\beta}_k \nonumber \\
  &=& \begin{pmatrix}
    1 & P_{k,1}(\xvec_{k,1}, \mOmega_{k}) & \cdots & P_{k,J_k}(\xvec_{k,1}, 
      \mOmega_{k}) \\
    \vdots & \vdots & \ddots & \vdots \\
    1 & P_{k,1}(\xvec_{nk}, \mOmega_{k}) & \cdots &
    P_{k,J_k}(\xvec_{nk}, \mOmega_{k})\end{pmatrix}
  \begin{pmatrix}
    \beta_{k,0} \\
    \beta_{k,1} \\
    \vdots \\
    \beta_{k,J_k}
  \end{pmatrix},
\end{eqnarray}
where the columns of $\mN_k(\mX_k, \mOmega_{k})$ are similarly defined
as in \eqref{eqn:designcolumn} but  now each distributional parameter $\eta_k$ is learned from the data through an AdaSoRT with distribution parameter specific set of nodes $\mathcal{N}_k=\mathcal{M}_k\,\dot\cup\, \mathcal{T}_k$, $\mathcal{M}_k=M_k$, $\mathcal{T}_k=T_k$ with $J_k+1=M_k+T_k+1$ columns.  Furthermore, $\mOmega_{k}$ is the set of weights associated with all paths of the $k$-th AdaSoRT and  $\betavec_{k}=(\beta_{k,0}, \beta_{k,1},\ldots,\beta_{k,J_k})^\top\in\dsR^{J_k}$ are the unknown parameters. 
Finally, we remark that in principal, for each distributional parameter $\theta_k$ a different feature matrix
$\mX_k$ can be used. To detail likelihood-based estimation in the following, we write 
$\betavec=(\betavec_{1},\ldots,\betavec_{K})^\top$ and 
$\mOmega=(\mOmega_{1},\ldots,\mOmega_{K})^\top$ for the vectors of all unknown parameters. 

\subsection{Estimation} \label{sec:estimation}

Next, we detail how estimation of the parameters $\boldsymbol{\beta}$ and $\omegavec$ of a  DAdaSoRT can be performed. Assume having a trained data set $\lbrace (\xvec_{i},y_i)\rbrace_{i=1,\ldots,n}$ of features $\xvec_i$ and outputs $y_i$ (w.l.o.g.~we write down the case for univariate $y$). Let $\mX_k=(\xvec_{1k}^\top,\ldots,\xvec_{nk}^\top)^\top$ for $k=1,\ldots,K$ be the feature matrix of $\theta_k$ and denote $\mX = (\mX_1^\top, \ldots, \mX_K^\top)^\top$ or the complete feature matrix of all $K$ parameters and all $n$ data points. The loss function is induced by the log-density $f_y$ chosen by the user and reads as 
\begin{equation*} \label{eqn:loglik}
\ell(\boldsymbol{\beta}, \omegavec ; \mathbf{y}, \mX) =
  \sum_{i = 1}^n \log \, f_y(y_i ; \theta_{i1} = h_1^{-1}(\eta_{1}(\xvec_{i1}; \boldsymbol{\beta}_{1}, \omegavec_{1})), \ldots,
  \theta_{iK} = h_K^{-1}(\eta_{K}(\xvec_{iK}; \boldsymbol{\beta}_{K}, \omegavec_{K}))). 
\end{equation*}
Denoting $t=0,1,2,\ldots,$ the outer iteration index and $k=1,\ldots, K$ the distributional parameter, our  algorithm then proceeds as follows: 

\begin{enumerate}
\item[\textbf{Step 1}] \textbf{Intialization of root nodes:} First, the $K$ root nodes or intercepts $\beta_{0k}$ are initialized, i.e., intercept only models are applied and we set
  $\eta_k^{(0)} = \beta_{0k}$ for $k = 1, \ldots, K$ 
\item[\textbf{Step 2}] \textbf{Splitting the root nodes and efficient sub-indexing:} Afterwards, the first
  soft split for each  $k$-th parameter is calculated using the current
  score vector $\mathbf{u}_k = \partial \ell(\boldsymbol{\beta}, \omegavec; \mathbf{y},
  \mX) / \partial \boldsymbol{\eta}_k$ and working weights\newline
  $\mathbf{W}_{kk} = -\mathrm{diag}(\partial^2 \ell(\boldsymbol{\beta}, \omegavec; \mathbf{y}, \mX) /
  \partial \boldsymbol{\eta}_k \partial \boldsymbol{\eta}_k^\top)$.
  Assume for this that we have a set of optimal weights
  $\omegavec_k = \{\omegavec_{1k}\}$ (estimated by e.g.~maximum likelihood), which determines the first
  split and results in the $n \times 2$ design matrices
  $\mN_{1k}(\mathbf{X}_k, \omegavec_k)$. At this point recall, that for one
  split only one vector of weights is needed. Next, we introduce the sub-index
   $c = 
  1, \ldots, \frac{J_k}{2}$ and denote by $\mN_{ck}( \cdot )$ the $n \times 2$ design matrices with two columns each, i.e.~$\mN_k(\mathbf{X}_k, \omegavec_k) =
  (\mathbf{1}, \mN_{1k}(\mathbf{X}_k, \omegavec_k), \ldots, \mN_{\frac{J_k}{2}k}(\mathbf{X}_k, \omegavec_k))$ for each $c$ and each $k$. Doing so, the estimation problem can be significantly simplified, since improving the model fit only
  requires $n \times 2$ matrices, matching the two-dimensional sub-vectors
   $\boldsymbol{\beta}_{ck}$ of $\boldsymbol{\beta}_k = (\beta_{0k},
  \boldsymbol{\beta}_{1k}^\top, \ldots, \boldsymbol{\beta}_{\frac{J_k}{2}k}^\top)^\top$. Now, given the first $n \times 2$ design matrices
  $\mN_{1k}( \cdot )$, the predictor is updated by first solving
  \begin{equation} \label{eqn:iwlsupdate}
  \boldsymbol{\beta}_{1k} =
    (\mN_{1k}(\mX_k, \omegavec_{k})^\top\mathbf{W}_{kk}\mN_{1k}(\mX_k, \omegavec_{k}) + \zeta\mathbf{I})^{-1}\mN_{1k}(\mX_k, \omegavec_{k})^\top\mathbf{u}_k
  \end{equation}
  and then setting $\boldsymbol{\eta}_k^{(t + 1)} = \boldsymbol{\eta}_k^{(t)} + \mN_{1k}(\mX_k, 
  \omegavec_{k})\boldsymbol{\beta}_{1k}$ (see, e.g., \citealp{Umlauf+Klein+Zeileis:2018}),
  where in the first split $t=1$. Here, the matrix $\zeta\mathbf{I}$
  is a ridge penalty matrix with very small values on the diagonal, e.g.\ $\zeta = 0.00001$,
  which only should ensure numerical stability.
\item[\textbf{Step 3}] \textbf{Building the trees:} In the next step, each column of $\mN_k(\mX_k, \omegavec_{k})$ is doubled as in Step
  2 when splitting the root node again assuming a set of weights that are now
  $\omegavec_k = \{\omegavec_{1k}, \omegavec_{2k}, \omegavec_{3k}\}$ for each $k$ (compare  Figure~\ref{fig:softtree}).
  For each of the  newly created design matrices, the parameters $\boldsymbol{\beta}_{2k}$ and $\boldsymbol{\beta}_{3k}$ are then calculated as in \eqref{eqn:iwlsupdate}, but in contrast to Step 2 the predictors are not updated
  immediately.  Instead, only the soft split in parameter $\theta_k$ that contributes the most to the log-likelihood is selected for
  updating. The same steps are performed for all distributional parameters $\theta_k$.
  \item[\textbf{Step 4}] \textbf{Updating the weights:}  Updating the weights $\omegavec_{rk}$, $r = 1, \ldots, S_k$, where $S_k$ is the number of weights at 
iteration $t$,  only requires first and second order derivatives of the log-likelihood w.r.t.\ $\omegavec_{rk}$,
which by the chain rule is easy to derive analytically. The weight updates can efficiently be computed  based on \eqref{eqn:iwlsupdate} using the current $\boldsymbol{\beta}_{ck}$. For instance, at iteration $t=1$, e.g.~for the first split, $\omegavec_{1k}$, is based on
$$
\text{ML}(\omegavec_{rk} | \yvec, \mX) = \underset{\omegavec_{rk}}{\text{arg max }} \ell(\omegavec_{rk} ; \yvec, \mX).
$$
  \end{enumerate}
Overall, this algorithm  successively improves the model fit by  calculating a new soft 
split at the ``best'' position for each $\theta_k$, while cycling through Steps 3, 4 for $t=1,2,\ldots,$ until convergence has been reached. We define the latter to be the case when no improvement in terms of the AIC can be achieved by further increasing the trees, for which the degrees of freedom are determined by the number of parameters in $\boldsymbol{\beta}$.



\pagebreak

\section{Extensions and Properties}\label{sec:properties}

\mbox{In this section we highlight  useful extensions of  DAdaSoRT and summarize selected  properties. }

\subsection{Avoiding Overfitting} \label{sec:overfitting}

\begin{figure}[!t]
\centering
\includegraphics[width=0.5\textwidth]{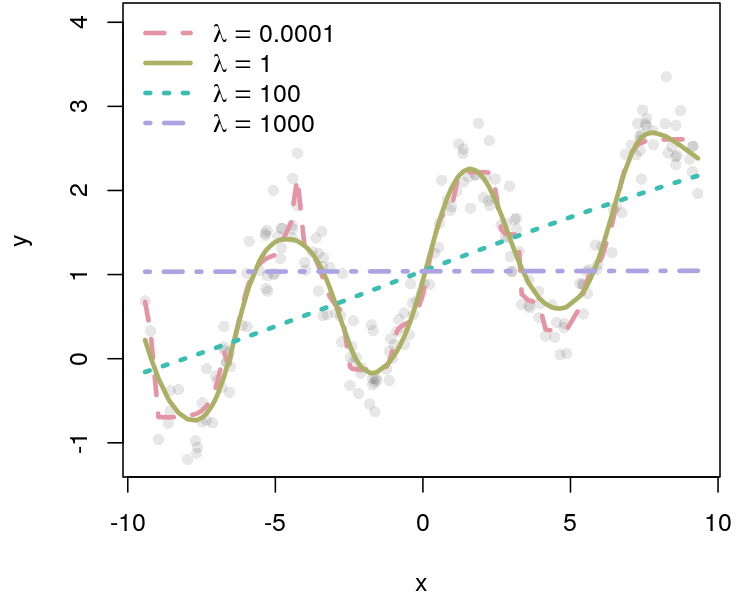}
\caption{\label{fig:penalty} Avoiding overfitting. Shrinkage effect of $\lambda$ using simulated data.}
\end{figure}
When maximizing the weights $\omegavec_{rk}$ an additional shrinkage penalty
$\lambda_k J(\omegavec_{rk}) = \lambda_k\omegavec_{rk}^\top\omegavec_{rk}$ can be imposed to avoid overfitting. This yields the  penalized maximum likelihood
$$
\text{pen ML}(\omegavec_{rk} | \yvec, \mX) = \underset{\omegavec_{rk}}{\text{arg max }} \ell(\omegavec_{rk} ; \yvec, \mX) - \lambda_k J(\omegavec_{rk}).
$$
The effect for different values of one shrinkage parameter $\lambda\equiv \lambda_{k}$ is shown if
Figure~\ref{fig:penalty}. For very small values of $\lambda$ the estimated function contains abrupt
jumps, e.g., at about $x = -4$. For larger values of $\lambda$ the fit gets smoother
and smoother until a linear function is estimated and in the limit a simple constant functional form would be fitted. The
selection of the shrinkage parameter is problem specific and is basically the only hyperparameter that needs 
to be tuned. In practice, we set $\lambda_k$ equal for all distributional parameters which usually
results in very good global model fits and accurate predictive distributions. Moreover, the determination of $\lambda_k$ is relatively simple:
First, $\lambda_k$ is set to a rather high value, so that the algorithm stops after a small number of iterations. Then $\lambda_k$ is decreased so that a steady state of the AIC or BIC can be achieved.
The reason for this is that too small values of $\lambda_k$ force the estimation algorithm to add
additional complexity to the estimated functions, see the jumps for $\lambda = 0.0001$ in Figure~\ref{fig:penalty}, leading
to overfitting of the final model. Overfitting is easily recognized by the fact that the AIC
continues to decrease with only very small improvements in the penalized likelihood. Of course this is a heuristic approach,
but in our experience it works very well, see the simulation Section~\ref{sec:simulation} and the
application Section~\ref{sec:application}.

\subsection{Effect Decomposition and Subsumed Special Cases}

Naturally, our DAdaSoRT allows for so-called effect decomposition of the predictors, which
can be useful in the GAMLSS context \citep[see, e.g.,][]{Kneib+Klein+Lang+Umlauf:2019} for interpretational purposes. For our case of a DAdaSoRT it could for instance be of interest to disentangle high-dimensional interactions through the soft tree predictor to increase predictive accuracy, while being able to interpret low-dimensional effects on specific aspects of the predictive distributions. For this purpose,
a structured part $\boldsymbol{\eta}_k^{\text{struct}}$ can be added to each predictor with
\begin{eqnarray*}
\boldsymbol{\eta}_k &=& \boldsymbol{\eta}_k^{\text{struct}} + \boldsymbol{\eta}_k^{\text{srt}} \\ \nonumber
  &=& \mathbf{X}\boldsymbol{\gamma}_k + \boldsymbol{\eta}_k^{\text{srt}}
\end{eqnarray*}
where predictor $\boldsymbol{\eta}_k^{\text{srt}}$ represents the AdaSoRT predictor and
is given in \eqref{eqn:addpred}; and $\mathbf{X}\boldsymbol{\gamma}_k$
are additional structured main effects (such as linear effects or smooth effects of univariate input features). In the simple case of linear effects only, i.e.~$\boldsymbol{\eta}_k^{\text{struct}}\equiv \boldsymbol{\eta}_k^{\text{lin}}$, this representation corresponds to
direct connectors for NN. However, more general representations
$$
\boldsymbol{\eta}_k^{\text{struct}} = f_{1k}(\xvec) + \ldots + f_{S_kk}(\xvec),
$$
with $s = 1, \ldots, S_k$ smooth effects of features, can be added to \eqref{eqn:addpred}.
The advantage of such an extension is the interpretability of the main effects, which is normally
not given for classical trees, forests and NNs. Therefore, identifiability constraints should be enforced as suggested by \citet{RueKolKle2021}.


\subsection{DAdaSo Forests}

Our DAdaSoRT naturally extends to ensemble of trees as in classical random forests \citep[e.g, by bagging][]{Breiman:1996, Yildiz2016}, which we refer to as DAdaSo forests.
We deliberately do not discuss the ensemble method very prominently, but focus on the already very good 
approximation properties of a single DAdaSoRT for brevity and readibility of our paper, but added them in 
both the simulation and the application sections. In the simulation Section~\ref{sec:simulation},
we found that DAdaSo forests have qualitatively similar or slightly better performance compared to a 
single DAdaSoRT for datasets of approximately 2000 observations and above. Below that, a single DAdaSoRT 
has slightly better performance, the reason for this is mainly due to the choice of the shrinkage
parameter in combination with bagging for forests, which we intentionally did not tune separately to 
emphasize the simplicity of the approach.

\subsection{Relation to Neural Network Universal Approximators}\label{sec:universalapproximator}

While it is is beyond the scope of the paper to rigorously investigate whether and in what sense soft trees are in general 
universal approximators, a valid starting point for future research can be the results on standard multilayer feedforward 
networks. For these, \citet{HorStiWhi1989} showed that they are capable of approximating any Borel measureable function on finite 
dimensional spaces provided sufficiently many hidden units are available. While the soft trees do not have hidden units, the 
splits, tree-depth and ensemble estimates of single trees combined with our choice for the mappings in \eqref{eqn:logistic} are 
expected to allow for a similar degree of flexibility with respect to its capability to approximate arbitrary complex functions.

\pagebreak

\section{Software Implementation} \label{sec:software}

DAdaSoRTs are implemented in the \proglang{R} package 
\pkg{softtrees} \citep{softtrees:R}. The package supports all families as implemented in the
\proglang{R} packags \pkg{bamlss} \citep{bamlss:R} and  \pkg{gamlss.dist}
\citep{gamlss.dist}. The \pkg{softtrees} package along all supplementary materials of
this article is available on GitHub: URL~\url{https://github.com/freezenik/softtrees}.
Further examples on how fit DAdaSoRT along with code are provided
in the manual pages of the package.

\section{Simulation Study} \label{sec:simulation}

In this section we provide empirical evidence of the efficacy of DAdaSoRT for several data generating processes compared to relevant competitors from the literature. To this end, we investigate the performance of DAdaSoRT with respect to bias 
using the root mean squared error ($\sqrt{\text{MSE}}$) of deviations from the true predictors $\eta_k$ (point accuracy), and overall predictive accuracy of predictive distributions used for 
probabilistic forecasting using proper scoring rules. Specifically, for continuous outputs we compute the continuous rank probability score \citep[CRPS;][]{Gneiting+Raftery:2007}, and discrete outputs the logarithmic score. 

\textbf{Simulation design}.
We simulate data sets of sizes $n=500,1000,5000,10000$ from three different distributions:
the normal distribution (\code{NO}), the Gumbel distribution (\code{GU}) and the
negative binomial distribution (\code{NBI}) from the \proglang{R} package
\pkg{gamlss.dist}. The package uses a specific naming convention for the parameters of the
distributions and supports up to four-parameter distributions. The parameters are
$\mu$, $\sigma$, $\nu$ and $\tau$. In the simulation study, we let the parameters $\mu$ and $\sigma$ 
depend on feature vectors. Since all the distributions studied in this framework have
two parameters, no specifications are required for $\nu$ and $\tau$. Specifically, we simulate data 
using 
$$
\eta_{\mu} = \left[\left(10\sin(\pi x_1x_2) + 20(x_3 - 0.5)^2  + 10 x_4 + 5 x_5\right) - 
  1.5\right]\frac{2}{26.48} + 1
$$for the predictor of parameter $\mu$
and
$$
\eta_{\sigma} = \left(\left(z_1^2 + \left(z_2z_3 - \frac{1}{z_2z_4}\right)^2\right)^{0.5} - 
  7.96\right)\frac{2}{1736.85} - 2.5,
$$
for $\sigma$ for all three data distributions. These predictors represent state-of-the art benchmark studies as the predictors are slightly scaled versions of the Friedman 1 and 2 functions \citep{Friedman:1991,Breiman:1996}. Following earlier work, the inputs  $x_1, \ldots, x_5$ and $z_1, \ldots, z_4$ are drawn independently  
from uniform distributions
with $x_q \sim U(0, 1)$, $q = 1, \ldots 4$, $z_1 \sim U(0, 100)$, $z_2 \sim U(40, 560\pi)$, $z_3 \sim U(0, 1)$ and
$z_4 \sim U(1, 11)$. Finally, for each of the settings, we replicate the simulation $100$ times.

\textbf{Benchmark methods}.
We compare the DAdaSoRT (denoted by \code{srt}) and the DAdaSo forests (denoted by \code{srf}) to a full Bayesian
structured additive distributional regression model,
estimated with the \proglang{R} package \pkg{bamlss} (denoted by 
\code{bamlss}) and distributional regression forests \citep{distforest:paper} as implemented in the 
\proglang{R} package \pkg{disttree} \citep[denoted by \code{distforest}]{distforest:R}.
Compared to \code{srt} 
representing a single tree, \code{distforest}s are  based on $1000$ trees each.
For \code{srf} we use only $100$ trees due to computation time.
For \code{bamlss} models, we use thin-plate splines \citep{Wood:2003} for nonlinear smooth effects, one 
for each covariate, estimated by Markov chain Monte Carlo simulation.

\begin{figure}[!t]
\centering
\includegraphics[width=0.8\textwidth]{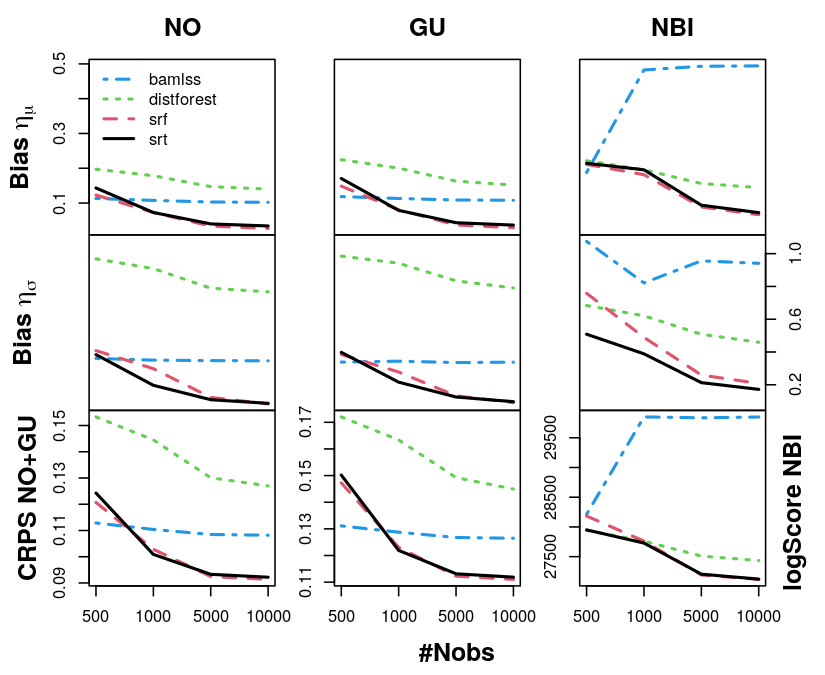}
\caption{\label{fig:sim_results} Simulation study.  The first two rows show the
  deviation from the true and the estimated predictor (Bias $\eta_k$) measured my the median
  $\sqrt{\text{MSE}}$ over the $100$ replications for different number of training samples and data
  distributions; the normal (\code{NO}), Gumbel (\code{GU}) and negative binomial (\code{NBI}). The true predictors $\eta_{\mu},\eta_{\sigma}$ are slightly scaled versions of  Friedman 1 and 2
  functions. The $\sqrt{\text{MSE}}\text{s}$ are calculated using a fixed data set
  with $10000$ test samples. The bottom row shows the corresponding CRPS for \code{NO} and \code{GU}
  and the logarithmic score for \code{NBI} using the same test data.}
\end{figure}

\textbf{Results}.
Figure~\ref{fig:sim_results} summarizes results for all methods and metrics. To achieve direct comparability of
results across the different settings, for each data distribution, we simulated one further test data set with $n=10000$ observations
that was fixed throughout. Shown results are the median metric across the 100 replications. 
Looking at the performance in terms of bias, 
CRPS and logarithmic scores, \code{srt} and \code{srf} clearly outperform the other methods.
Only for very small 
data sets with $500$ observations the \code{bamlss} models show slightly better results for distribution 
\code{NO} and \code{GU}.
Furthermore, \code{srf} is slightly better in all settings according the predictor for parameter $\mu$
and is slightly better according the CRPS and logarithmic score for $n > 1000$.
For $n \leq 1000$ \code{srt} is slightly better according the predictor for  $\sigma$ for \code{NO}
and \code{GU} and for all $n$ for \code{NBI} and also according to the logarithmic score
for $500$ training points for \code{NBI}.
The reason for this is that \code{srf} is computed using bagging and we did not adapt the
shrinkage parameter separately to emphasize the simplicity of the approach, i.e.,
\code{srf} can most likely outperform more settings.
Interestingly, the \code{distforest} seems to have the worst 
performance for the distributions \code{NO} and \code{GU} for all sizes of training data, i.e., the method apparently cannot find the 
non-linear structure of the simulated predictors very well. Only in the count data setting with
$500$ and $1000$ training points the results of the logarithmic score for \code{distforest} are very
similar to \code{srt} and \code{srf}.

\textbf{Conclusion}.
In summary, it can be clearly identified that  \code{srt} and \code{srf} are very well
suited to finding non-linear relationships in the data compared to
the conventional \code{distforest}. As expected, the \code{bamlss} models
cannot approximate the non-linear structures very well. In contrast, both DAdaSoRT and DAdaSo forests
are very robust across settings and are an attractive alternative to the considered 
benchmark methods in particular with large data and complex relations between outputs and inputs.
The results for \code{distforest} and \code{bamlss} were calculated with the defaults of the software 
packages, only the number of trees was increased from 500 to 1000 for the \code{distforest}s. 
The defaults for \code{distforest} are set so that very flexible structures can be found in principle.
The only tuning parameter for DAdaSoRT is the shrinkage penalty for the 
weights, see Section~\ref{sec:estimation}, for which we have simply identified reasonably good values
with the strategy as described in Section~\ref{sec:overfitting}, meaning that further tuning
could even improve the results, especially for the DAdaSo forests.

\section{Solar Cycle Forecasting} \label{sec:application}

The Sun's activity has been closely observed and recorded for a very long time. The solar activity is 
determined by numerous explosions, also known as coronal mass ejections and solar flares, which emit a 
tremendous amount of energy. Solar flares emit X-rays and magnetic fields that can literally bombard 
earth as geomagnetic storms and thus affect activity on earth \citep[see, e.g.;][]{Hathaway2015}. For instance, strong solar activity can change polarity of satellites and  damage its electronics.
Typically, these eruptions occur near sunspots, i.e., 
the more sunspots visible on the solar surface, the greater the solar activity. The number of sunspots 
follows a certain periodicity with a cycle of about eleven years. Therefore,  the
number of sunspots and  the prediction of solar cycles are of particular interest and have thus been recorded
since 1749. In the past decades, 
various models have been developed to forecast the number of sunspots. For example, in one of the most 
recent articles on sunspots at the time of writing \citep{Dang2022}, non-deep learning methods are compared with deep 
 ensemble learning methods for prediction.
The \emph{National Aeronautics and Space Administration} (NASA) solar cycle forecast is available at
\url{https://www.nasa.gov/msfcsolar/} (last viewed on 2022/08/11) from the
\emph{Space Environments Team in the Natural Environments Branch of the Engineering Directorate at Marshall Space Flight Center (MSFC)}.
According to MSFC, monthly smoothed sunspot series are used to construct forecasts  based on a regression model
estimating the deviation from a mean solar cycle, also called MSFC solar activity future estimation (MSAFE) model \citep{Suggs2017TheMS}.
The model is updated monthly, and the first published forecasts began in March 1999, and to our
knowledge the model has not been changed since then.

In most studies, the different methods are typically compared using point estimates and then
evaluated using, e.g., the MSE. To contribute to the topic of sunspot forecasting we use our
proposed \code{srt} from Section~\ref{sec:estimation}) and benchmark it against point forecasts from
the NASA's MSAFE model (\code{NASA}) and the distributional regression forest (\code{distforest}) using 
the MSE. 
In addition, unlike other studies, we assess the full
probabilistic predictive ability for \code{srt} and \code{distforest} using the CRPS score.

Our analysis is based on monthly mean sunspot data from November 1833 to June 2022 for the response $y_t$ with $t = 1, \ldots, 2263$,
which is provided by \citet{sidc}.
\begin{figure}[!t]
\centering
\includegraphics[width=1\textwidth]{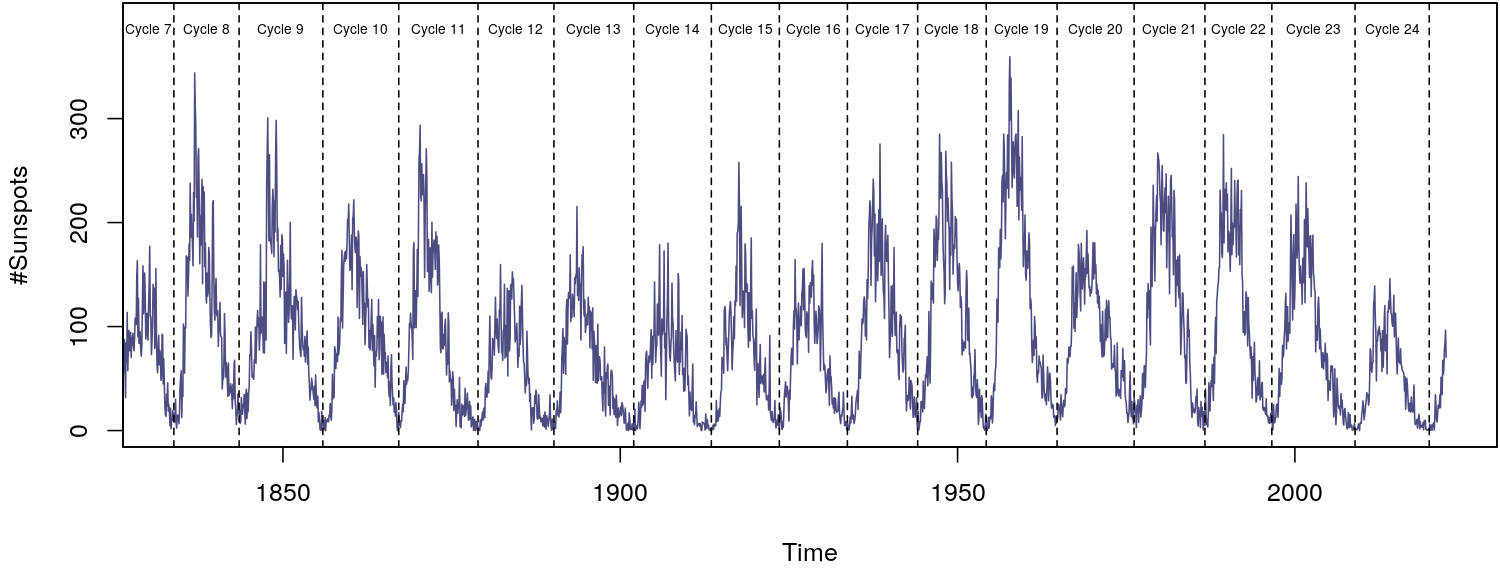}
\caption{\label{fig:sunspot_data} Solar cylce forecasting. Monthly mean sunspot data  from November
  1833 to June 2022.}
\end{figure}
The data $y_t$ is shown in Figure~\ref{fig:sunspot_data}.
To evaluate forecast performance, we use all 277 available \code{NASA} forecasts,
with the first forecast starting in March 1999 and the last in September 2021, and compare them to the 
forecasts obtained by \code{srt} and \code{distforest}. For training  we  only use
data before the respective \code{NASA} forecast starts. More precisely, we use lagged data of sunspot 
numbers as inputs for \code{srt} and \code{distforest} (24 monthly lags, $y_{t-1}, y_{t-2},
\ldots, y_{t-24}$, plus annual lags up to 35 years into the past,
$y_{t - 12 \cdot j}$, $j = 3, \ldots, 35$)
and compute predictions by recursive multi-step forecasting \citep[see, e.g.;][] {Taieb2014}. We additionally
transform the response with $\tilde{y}_t = \sqrt{y_t + 0.001}$ to achieve better numerical stability.
In total, we use three different distributions (implemented in \pkg{gamlss.dist}) for comparison,
the normal distribution (\code{NO}),
the gamma distribution (\code{GA}) and the t family distribution (\code{TF}).

Because of the lagged input data structure, the choice of the shrinkage parameter $\lambda$
used in \code{srt} is more likely to influence the model performance,
more than in usual regression settings. Therefore,
we tested \code{srt} with $\boldsymbol{\lambda} = (1, 5, 10, 50, 100, 500, 1000)^\top$ and used the shrinkage
parameter with the best forecast performance. In this way we manage to get a sense of different
shrinkage and forecast horizons, the shortest horizon being nine months and the longest to date 236 months.
For a detailed analysis, we evaluated \code{short}, \code{medium} and \code{long} forecast horizons (0 to 99 months, 100 to 199 months and $\geq$ 200 months, respectively).
\begin{table}
\centering
\begin{tabular}{llrrr}
 & Horizon & 1st & 2nd & 3rd \\ \hline
\multirow{3}{*}{CRPS} & short & \code{srt-TF}: \textbf{0.7835} & \code{srt-NO}: 0.7870 & \code{srt-GA}: 0.8678\\
 & medium & \code{srt-NO}: \textbf{0.9325} & \code{srt-GA}: 0.9679 & \code{srt-TF}: 0.9739\\
 & long & \code{srt-GA}: \textbf{1.0620} & \code{distforest-GA}:  1.0700 & \code{distforest-TF}: 1.0710 \\\hline
\multirow{3}{*}{MSE} & short & \code{srt-NO}: \textbf{147.50} & \code{srt-TF}: 150.80 & \code{NASA}: 288.50\\
 & medium & \code{srt-GA}: \textbf{409.50} & \code{srt-NO}: 435.70 & \code{distforest-TF}: 439.80\\
 & long & \code{srt-GA}: \textbf{562.50} & \code{distforest-GA}: 590.90 & \code{distforest-NO}: 626.80 \\\hline
\end{tabular}
\caption{\label{tab:sunspots} Solar cycle forecasting. Median CRPS and MSE scores for
  different forecast horizons, models and distributions, second and third best, column 1st, 2nd and 3rd  respectively.}
\end{table}
The results are shown in Table~\ref{tab:sunspots} and indicate that our proposed \code{srt} method
outperforms \code{distforest} and also the forecast from \code{NASA} over all forecast horizons.
The choice of the best distribution  changes with the length of the forecast horizon indicating
that full probabilistic models have advantages compared to models  for the mean only.
In particular, for long-term forecasts, \code{srt} in combination with the \code{GA} distribution seems to
significantly improve the prediction quality compared to \code{distforest} and \code{NASA},
according to the reported metrics.
\begin{figure}[!t]
\centering
\includegraphics[width=1\textwidth]{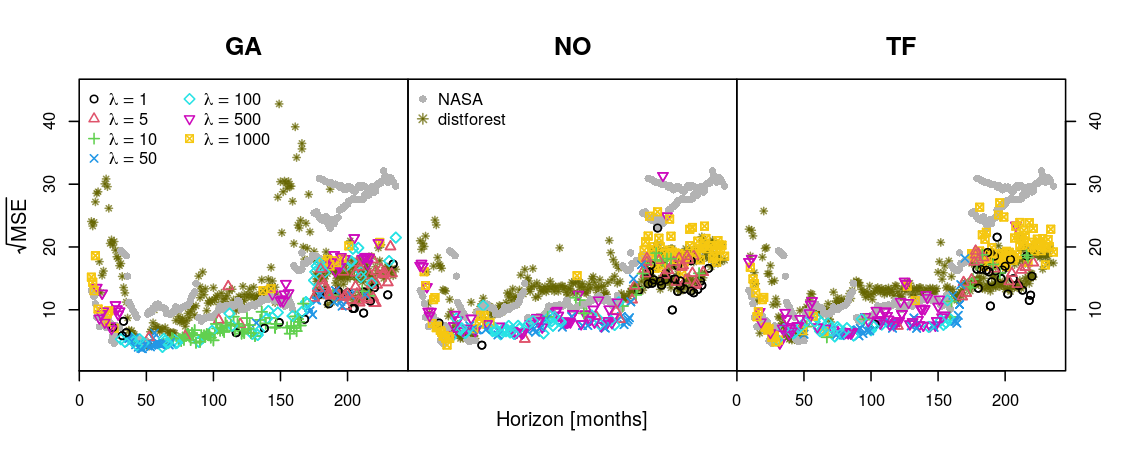}
\caption{\label{fig:lambdas} Solar cylce forecasting. Shown are the median  $\sqrt{\text{MSE}}$s
(y-axis) across the forecast horizons in months (x-axis)  for the \code{NO} (left), \code{GA} (middle) and \code{TF} (right) distributions. The different colors and shapes correspond to the different values for $\lambda$. In addition, corresponding MSEs from \code{NASA} and \code{distforest} are indicated
  by the gray filled dots and olive asterix, respectively.}
\end{figure}
In Figure~\ref{fig:lambdas} the results for MSEs over the different forecast horizons are shown in more detail and
illustrates that the \code{srt} method has a very good performance, especially for long term forecasts. Only for very short forecasts, up to 25 months,
\code{NASA} seems to be slightly superior compared to \code{srt}.  Overall, \code{distforest} performs worst in particular for short-term predictions. Note that the best distribution for \code{srt}, \code{GA}, is leading to very high values
in the MSEs for \code{distforest}, while \code{srt} shows a stable performance for all distributions.
In addition, Figure~\ref{fig:lambdas} shows that the choice of $\lambda$ for \code{srt} for long-term
forecasts using the best-performing \code{GA} distribution consistently tends toward rather small values,
implying a higher degree of model input interactions, and higher values with less interaction
for short-term forecasts.

\begin{figure}[!t]
\centering
\includegraphics[width=1\textwidth]{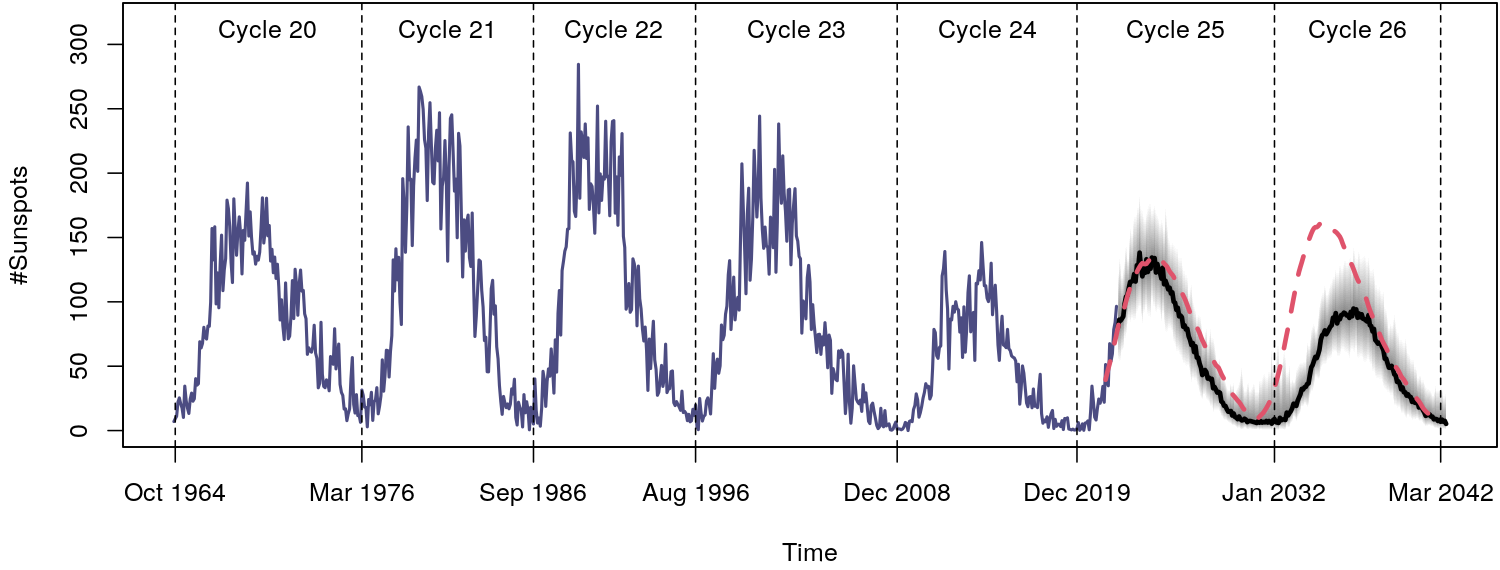}
\caption{\label{fig:sunspot_forecast} Solar cycle forecasting. Shown are monthly sunspot observations (dark blue lines) together with forecasts
  for solar cycle 25 and 26. The black line represents forecasts from the ensemble of 2000 \code{srt} trees with the \code{GA} distribution 
  including 5\% and 95\% prediction intervals shown by the gray shaded area. The red dashed line represents
  the forecast from \code{NASA}.}
\end{figure}
As commonly the long term forecasts get the most attention, we finally estimate a model with all available data to try
to predict the solar cycle 25 and 26, which corresponds to a \code{long} forecast of 242 months. To further minimize
the uncertainties of the forecast we create a DAdaSo forest, \code{srf}, for the \code{GA} distribution.
To do so, we train a total of 2000 \code{srt}
trees on 63\% of the data (bagging) and with $\lambda = 2$ for each tree. The final ensemble forecast for solar cycle 25 and 26 is
shown in Figure~\ref{fig:sunspot_forecast}. It can be seen that the prediction for solar cycle 25 is essentially the same as provided by
\code{NASA}, which consolidates the fact that \code{NASA}'s prediction could more or less come true in this way.
For solar cycle 26 the differences in amplitude are clear, the \code{srf} predicts a much weaker cycle than
\code{NASA}, while the predicted cycle of \code{srf} starts later.
The shown prediction intervals of \code{srf}
show a larger uncertainty especially in the maximum of the cycles and further illustrate the large difference in
the forecast for solar cycle 26. At first sight, the shape of our predicted
cycle 26 seems rather unrealistic compared to \code{NASA}'s forecast, but on closer inspection,
solar cycles 7, 12, 13 and 14 (and earlier 5 and 6, not shown in Figure~\ref{fig:sunspot_data}) are quite similar to our prediction.

To conclude, the proposed \code{srt} and \code{srf} can be seen as competitive methods for sunspot 
prediction. Especially the fact that with \code{srt} a full probabilistic forecast can be computed
includes significant advantages for estimating the forecast uncertainty compared to point predictions.
The predictive performance of the computed \code{long}-term forecast will be seen in the future,
but we believe that \code{srt} and \code{srf} can certainly be a good alternative,
although of course more work needs to done for better model specification and the
definition of the hyperparameters.

\section{Summary} \label{sec:summary}

In this paper, we developed SoRT for the distributional learning setting and proposed a number of
extensions, including ensembles of DAdaSoRT. DAdaSoRT are a very flexible class of models that can be
used for full probabilistic forecasting. For efficient estimation we present an adaptive algorithm,
where the size of the SoRT is determined by information criteria such as AIC or BIC, leading to models 
with relatively few degrees of freedom. We show in an extensive simulation study that the performance of 
DAdaSoRT is better than, or at least similar to, that of a distributional random forest. The simulation 
results are also reflected in the solar activity prediction application.
In summary, the proposed distributional adaptive soft regression tree (DAdaSoRT) is very competitive, 
especially in cases with a high degree of interactions amongst input features. However,
in future research, some improvements need to be made in terms of algorithmic performance
when using large datasets as well as the discussed extensions by decomposing the predictor
into main and interaction effects. In addition, the question of the (universal) approximation 
capabilities remains open. We also plan to extend DAdaSoRT in the Bayesian context
to obtain full Bayesian inference.

\section*{Acknowledgments}

\begin{leftbar}
Nikolaus Umlauf was supported by the Austrian Science Fund (FWF) grant number 33941.
Nadja Klein was supported by the Deutsche Forschungsgemeinschaft (DFG, German Research Foundation)
through the Emmy Noether grant KL 3037/1-1.
\end{leftbar}


\bibliography{softtrees}


\end{document}